\documentclass[preprint]{vgtc}               

\ifpdf
  \pdfoutput=1\relax                
  \usepackage{graphicx}                
  \DeclareGraphicsExtensions{.pdf,.png,.jpg,.jpeg} 
\else
  \usepackage{graphicx}                
  \DeclareGraphicsExtensions{.eps}     
\fi%

\graphicspath{{figures/}{pictures/}{images/}{./}} 

\usepackage{microtype}                 
\PassOptionsToPackage{warn}{textcomp}  
\usepackage{textcomp}                  
\usepackage{mathptmx}                  
\usepackage{times}                     
\usepackage{cite}                      
\usepackage{tabu}                      
\usepackage{booktabs}                  

\usepackage{paralist}
\usepackage{tikz}
\definecolor{label-background}{HTML}{eeee77}
\tikzset{My Style/.style={ font=\sffamily\large, red, draw=blue, fill=yellow!20, minimum size=0.5cm}}
\tikzset{
  image label/.style={
      circle,
      fill=label-background,
      draw=black,
      opacity=0.75,
      inner sep=1.25pt,
      font=\sffamily\large, 
      anchor= west, 
      at={(0,0.5)}
    }
}

\onlineid{0}

\vgtccategory{Research}

\vgtcinsertpkg

\preprinttext{\parbox{1\textwidth}{$\copyright$ 2022~IEEE. This is the author's version of the article that has been published in the proceedings of the 2022 IEEE Workshop on TRust and EXpertise in Visual Analytics (TREX).
The final version of this record is available at: \href{https://doi.org/10.1109/TREX57753.2022.00006}{\color{blue}10.1109/TREX57753.2022.00006}.}}

\ieeedoi{10.1109/TREX57753.2022.00006}

\title{Trustworthy Visual Analytics in Clinical Gait Analysis: \\ A Case Study for Patients with Cerebral Palsy}

\author{Alexander Rind\thanks{Alexander Rind and Djordje Slijepčević equally contributed to this paper and are both to be regarded as first authors.} \thanks{e-mail: \{firstname\}.\{lastname\}@fhstp.ac.at}\\ %
        \scriptsize Inst.\ of Creative\textbackslash{}Media$/$Technologies\\ \scriptsize St.~P\"olten Univ.\ of Applied Sciences \\ \scriptsize \quad%
\and Djordje Slijepčević$^{*\dagger}$\\ %
        \scriptsize Inst.\ of Creative\textbackslash{}Media$/$Technologies\\ \scriptsize St.~P\"olten Univ.\  of Applied Sciences %
\and Matthias Zeppelzauer$^\dagger$\\ %
        \scriptsize Inst.\ of Creative\textbackslash{}Media$/$Technologies\\ \scriptsize St.~P\"olten Univ.\  of Applied Sciences %
\and  Fabian Unglaube\thanks{e-mail: \{firstname\}.\{lastname\}@oss.at}\\ %
     \scriptsize Orthopaedic Hospital Vienna-Speising %
\and Andreas Kranzl$^\ddag $\\ %
     \scriptsize Orthopaedic Hospital Vienna-Speising %
\and Brian Horsak$^\dagger$\\ %
        \scriptsize  CDHSI \& Inst.\ of Health Sciences\\ \scriptsize  St.~P\"olten Univ.\  of Applied Sciences %
}

\abstract{%
Three-dimensional clinical gait analysis is essential for selecting optimal treatment interventions for patients with cerebral palsy (CP), but generates a large amount of time series data. For the automated analysis of these data, machine learning approaches yield promising results. However, due to their black-box nature, such approaches are often mistrusted by clinicians.
We propose gaitXplorer, a visual analytics approach for the classification of CP-related gait patterns that integrates Grad-CAM, a well-established explainable artificial intelligence algorithm, for explanations of machine learning classifications.  
Regions of high relevance for classification are  highlighted in the interactive visual interface.
The approach is evaluated in a case study with two clinical gait experts.
They inspected the explanations for a sample of eight patients using the visual interface and expressed which relevance scores they found trustworthy and which they found suspicious. 
Overall, the clinicians gave positive feedback on the approach as it allowed them a better understanding of which regions in the data were relevant for the classification.%
} 

\CCScatlist{
  \CCScatTwelve{Human-centered computing}{Visu\-al\-iza\-tion}{Visualization application domains}{Visual analytics};
  \CCScatTwelve{Computing methodologies}{Machine learning}{}{};
  \CCScatTwelve{Applied computing}{Life and medical sciences}{}{}

}

\begin{document}

\firstsection{Introduction}

\maketitle
Impairments in our ability to walk pose a major threat to participation in social activities and the labor market, and are therefore closely linked to quality of life. 
In children, one of the most common causes of physical disability is cerebral palsy (CP). It is diagnosed approximately in 2.5 per 1,000 births in developed countries~\cite{mcintyre_continually_2018} and constitutes a group of neurological disorders associated with varying symptoms such as tremors, muscle weakness, muscle stiffness, and spasticity~\cite{perezcerebral}. These symptoms affect the child's motor functions, including the ability to walk. The most common causes for CP are brain lesions that occur shortly before or after birth. These lesions can result in musculoskeletal impairments, which develop and worsen throughout childhood and adolescence~\cite{graham_cerebral_2016}.
To provide the best possible care for children with CP, a comprehensive \textit{physical examination and quantification of the patients’ gait pattern} have become essential tools in the treatment of CP. \textit{Three-dimensional gait analysis} is the gold standard for quantifying human movement and determining how musculoskeletal impairments affect a child's ability to walk, both in clinical and research settings~\cite{wren_efficacy_2011}. 

Clinical gait analyses produce a vast amount of data with high-dimensionality, temporal dependencies, strong variability, non-linear relationships, and inter-correlations among variables~\cite{chau_review_2001}. 
This is especially true for patients with CP, who often exhibit high variability in their gait pattern, which can significantly complicate the interpretation of clinical gait analysis data~\cite{liptak_health_2004}. The diversity in gait patterns and clinical representations has led researchers in the past to develop automated gait classification methods that can aid in diagnosis, clinical decision making, and communication~\cite{papageorgiou_systematic_2019}. 
These methods enable the classification of gait patterns into clinically meaningful groups, which can be distinguished from each other based on a set of defined (biomechanical) variables~\cite{dobson_gait_2007}. One of the most frequently used clinical classification schemes for CP is the one proposed by Rodda et al.~\cite{rodda_classification_2001}. Briefly explained, it allows to distinguish CP-related gait patterns in the sagittal plane such as drop foot, true equinus, jump knee, apparent equinus, and crouch gait depending on whether one or both sides of the body are affected (i.e., spastic hemiplegia vs. spastic diplegia). Accurate and objective classification of these patterns is critical, as this information is the basis on which clinicians make decisions about optimal treatment interventions. To support clinicians in diagnosis of CP-related gait patterns, great efforts have been undertaken in the past to develop automated classification algorithms~\cite{papageorgiou_systematic_2019,chia_decision_2020-1,darbandi_automatic_2020,sangeux_sagittal_2015, zhang2019application, ferrari2019gait}. 
While these methods are promising for supporting clinicians in their daily routines, most of the developed approaches, however, employ complex classification methods which have a black-box character, making it impossible to understand why the automated algorithm made a particular classification. 

Clinical experts have extensive experience with CP-related gait patterns from patient observation and the diagnostic literature, but they cannot effectively explain the functioning of a given black-box model. 
Transparency is, however, essential in the medical field.
Its absence leads to a lack of trust in the algorithm, as it is not clear whether the model makes its predictions based on clinically relevant features or based on bias in the data (e.g., due to walking speed differences, marker misplacement, or other data collection issues). 
Thus, automated classification based on machine learning (ML) has not been widely used in clinical practice.
To overcome this problem, this work combines explainable artificial intelligence (XAI) \cite{adadi2018peeking, holzinger2017we} with visual analytics \cite{andrienko_2018_viewing, keim_2010_mastering, thomas_2005_illuminating}.

We present \textit{gaitXplorer,} an explainability-enriched visual analytics approach for the classification of gait patterns, and demonstrate it in a case study with gait data of children with CP.  
We exploit the potential of modern ML techniques in combination with visualizations of clinical time series.
To make the trained ML models more transparent, our approach highlights the regions in the patients' gait data that are particularly relevant for the prediction of the model. In addition, our approach allows for the comparison of patient time series data to a variety of subgroups in the dataset, allowing clinicians to examine the entirety of their dataset that is used to train ML models.
The developed visual analytics approach enables clinical experts to engage in joint human-machine reasoning and to compare the grounding of the automated classification methods with their expert knowledge. 
We conducted this case study in the course of a long-term collaboration with clinical gait experts from the Orthopaedic Hospital Vienna-Speising in Austria. 
It provides a unique opportunity to study explainability-enriched visual analytics approaches in a real-world setting and explore the effects it has on the trust of clinical experts in automated gait classification.

This work is structured as follows:
After the Related Work in \autoref{sec:relwork}, we characterize and abstract the application domain of clinical gait analysis in \autoref{sec:domain}. 
In \autoref{sec:design}, we describe our visual analytics approach and its underlying design decisions. 
\autoref{sec:eval} reports on the evaluation of the case study with our clinical collaboration partners on gait data from eight patients with CP. 
\autoref{sec:discuss} discusses the clinicians' feedback, limitations, and future work and 
\autoref{sec:conclusion} provides a brief summary and conclusion of the presented work.

\begin{figure}[t]
\begin{center}
    \includegraphics[width=1\linewidth]{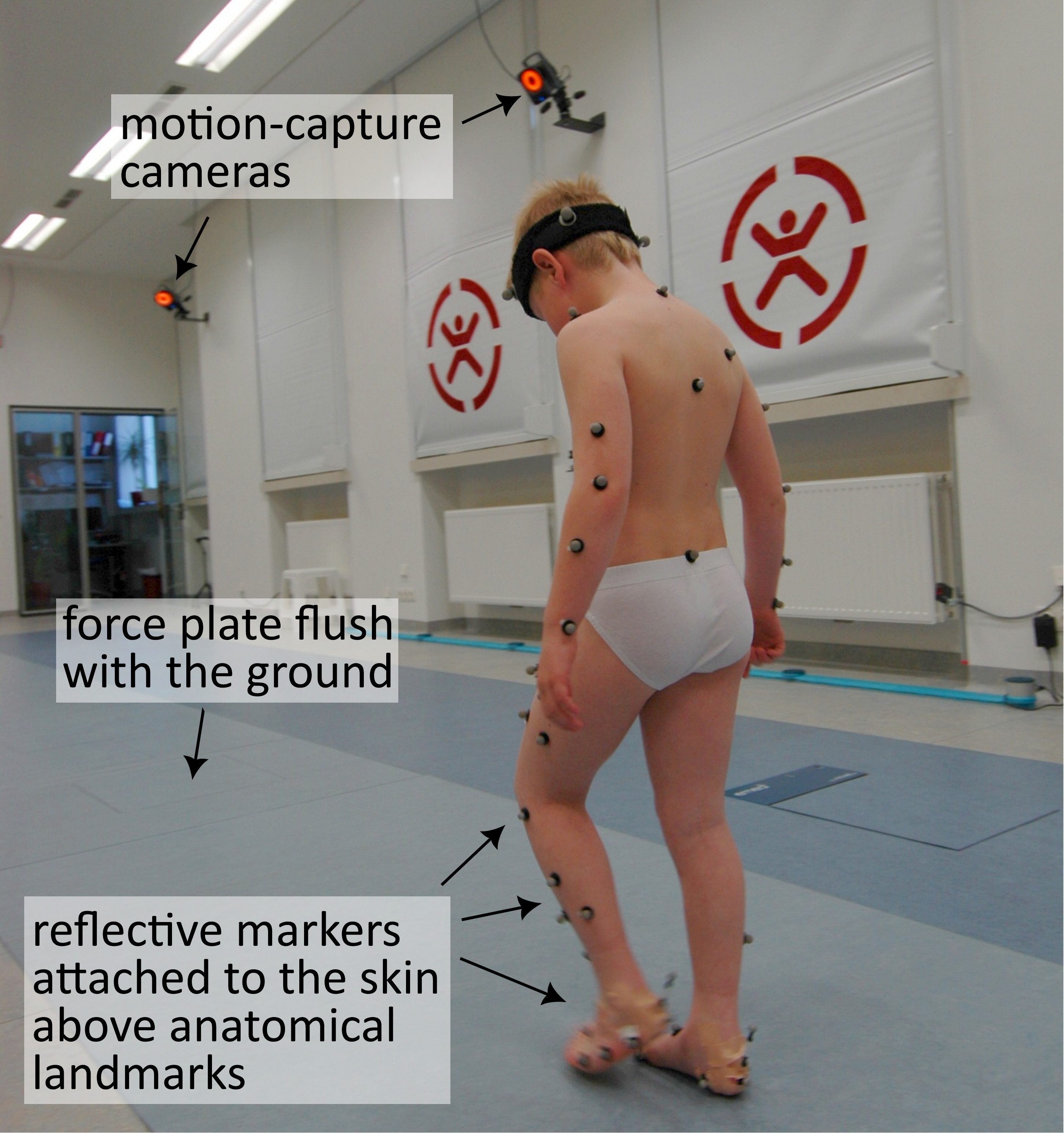}
\end{center}
\caption{A typical gait analysis scenario. The patient is equipped with a marker set and walks up and down an approximately 10-meter walkway. A set of motion-capture cameras tracks the three-dimensional trajectories of the markers and force plates embedded flush with the floor measure the external ground reaction forces (GRFs). \newline (Photo \copyright{}~Orthopaedic Hospital Vienna-Speising)}
\label{fig:3dgaLab}
\end{figure}

\section{Related Work}
\label{sec:relwork}

Due to the ability to analyze a large amount of gait data in a cost-effective, fast, and objective manner, there is a growing interest in the use of ML in the field of gait analysis~\cite{figueiredo2018automatic, halilaj2018machine}. ML methods have been successfully applied to patients with, for example, stroke~\cite{cui2018simultaneous}, Parkinson's disease~\cite{wahid2015classification}, osteoarthritis~\cite{long2017predicting}, or various functional gait disorders~\cite{slijepcevic2017automatic, slijepcevic2020input}. Automatic classification of CP-related gait patterns, in particular, is a frequently addressed topic in the literature~\cite{papageorgiou_systematic_2019}. Ferrari et al.~\cite{ferrari2019gait} compared multilayer perceptrons (MLP), support vector machines (SVM), and recurrent neural networks (RNNs) in a classification task with four self-defined CP-related gait patterns, and the latter showed the best performance. Zhang and Ma~\cite{zhang2019application} examined seven machine learning algorithms (e.g., MLP, SVM, Random Forest (RF)) for the classification of CP-related gait patterns as defined by Rodda et al.~\cite{rodda_classification_2001}, with MLPs performing best. For the same task, Darbandi et al.~\cite{darbandi_automatic_2020} used a fuzzy algorithm to translate the expert knowledge into rules and to perform fuzzy clustering. Sangeux, Rodda, and Graham~\cite{sangeux_sagittal_2015} presented the plantarflexor-knee extension index, a non-ML-based approach for the same classification task. The index represents the distance of the patient's ankle and knee kinematics in mid-stance from normative data~\cite{sangeux_sagittal_2015}. 
Chia et al.~\cite{chia_decision_2020-1} proposed a decision support system consisting of two models using physical examination and kinematic data to identify CP-related impairments and surgical recommendations. They employed a stratified RF and leveraged feature importance to provide an explanation for the prediction.

Besides being relatively accurate, complex ML models have a limitation, namely their black-box character~\cite{adadi2018peeking}. Thus, it is difficult to determine what an ML model has learned from the data or why certain decisions are made. As a result, even well-functioning ML models are rarely used in clinical practice~\cite{holzinger2017we}. 

To overcome this problem, Slijepčević, Horst et al.~\cite{slijepcevic2021explaining} proposed several XAI approaches based on Layer-wise Relevance Propagation~(LRP)~\cite{bach2015pixel} to explain the functioning of ML models trained to classify different functional gait disorders. Dindorf et al.~\cite{dindorf2020interpretability} utilized Local Interpretable Model-Agnostic Explanations~(LIME)~\cite{ribeiro2016should} to explain an ML model distinguishing between healthy controls and patients after total hip arthroplasty.

Interestingly, in the field of visualization and visual analytics, there is a relatively small body of literature that is focused on clinical gait analysis and related topics. 
In particular, the KAVAGait~\cite{wagner_2018_kavagait} approach supports exploring gait data in a clinical setting and interactively externalizing expert knowledge about gait pattern classifications. However, it utilizes only spatiotemporal parameters (e.g., walking speed).
Three approaches visualize motion data of the upper extremities for rehabilitation:
The visualization approach by Krekel et al.~\cite{krekel_2010_visual} shows kinematic data of shoulder and arm joints in a combination of three-dimensional views and time series plots.
The Motion Browser \cite{chan_2019_motiona} provides a time series visualization to analyze muscle activity patterns in an approach that integrates it with motion data and video information.
The NE-Motion~\cite{contreras_2021_nemotion} approach supports assessing movement impairment and compensation of the upper extremities caused by stroke. It visualizes the relationships between joint angle data based on a graph learning method.
In a veterinary setting, the FuryExplorer \cite{wilhelm_furyexplorer_2015} approach visualizes motion capture data of horses for analysis in lameness recovery.  
Several visual analytics approaches, i.e.,  MotionExplorer~\cite{bernard_motionexplorer_2013}, 
GestureAnalyzer \cite{jang_2014_gestureanalyzer}, and 
MotionFlow~\cite{jang_2015_motionflow}, support exploratory search for motion sequences in a hierarchically clustered motion tracking database. 
Bernard et al.~\cite{bernard_2017_visualinteractive} also presented a visual-interactive approach for the semi-supervised labeling of human motion capture data. 
Thus, we could not identify any previous work on visual analytics with three-dimensional motion capture data collected in a clinical gait laboratory. 

\section{Problem Characterization and Abstraction}
\label{sec:domain}

In clinical and research settings, gait performance is most frequently quantified using an opto-electronic motion capture system~(Figure~\ref{fig:3dgaLab}). These are very similar to the hardware used in the film and animation industry. A set of retro-reflective spherical markers are attached to the skin above bony landmarks. The three-dimensional coordinates of these markers are tracked by the motion capture system. Time-synchronized with these trajectories, external ground reaction forces (GRFs) are measured using a set of force plates that are flush with the floor. These data can then be used to calculate variables such as kinematics (e.g., knee flexion-extension angle), kinetics (e.g., knee flexion moment), and spatiotemporal parameters (e.g., walking speed, step length, step time) by utilizing a biomechanical model and inverse dynamic calculations. Most of the kinematic and kinetic data are calculated for all three anatomical planes. To allow comparison in the time domain to other patients or to normative data, the time series in gait analysis are time-normalized to one gait cycle (0-100\% gait cycle), defined as the time interval between successive initial foot contacts of the same foot with the ground. The data are summarized in a visual gait report and used to inform clinicians during decision making. 

For this case study, we focus on the time series data for the lower body. In combination with spatiotemporal parameters and various indicators of gait quality, the time series form the most important basis for diagnosis. The clinical experts work with a total of 58 time series (kinematics and kinetics). 
These consist of 
\begin{compactitem}
    \item 
the joint angles (degrees) of the pelvis, hip, knee, and ankle in all three anatomical planes ($n=12$), 
    \item 
the foot progression angle and the angle between the floor and the bottom of the foot ($n=2$), 
    \item 
the joint moments (Nm/kg) of the hip, knee, and ankle in three anatomical planes ($n=9$), 
    \item 
the power (W/kg) for the hip, knee, and ankle joints ($n=3$), 
    \item 
and the GRF (N or \% body weight) in three anatomical planes ($n=3$)
\end{compactitem}
for the left and right sides ($29 * 2 = 58$).
While the time series are measured in different units and exhibit diverging value ranges, they still follow typical patterns which are meaningful to clinicians.

The primary task of the clinical experts involved in the present case study is medical decision making. For this purpose accurate and objective classification of gait patterns is a prerequisite to offer optimal treatment strategies to the patient. 
To assist them in the process of data analysis, there is a strong interest in integrating automated analysis algorithms into their workflows.
The clinical experts have annotated a large training dataset based on anonymized patient records and are collaborating with us to explore the utility of ML models to classify gait patterns into clinically meaningful categories.  
For both scenarios, medical decision making and ML model evaluation, the clinical experts need a visual analytics system that 
\begin{compactitem}
    \item 
shows the patient's biomechanical data in a familiar and efficient way,
    \item 
allows comparison of the patient's biomechanical data with the data of specific patient groups,
    \item 
automatically suggests a gait pattern classification, and 
    \item 
provides a detailed explanation for this classification in terms of the input gait measurement data.
\end{compactitem}

\section{Design and Implementation}
\label{sec:design}

Our prototypical visual analytics approach \textit{gaitXplorer} provides access to a list of patients that have been newly classified and should be inspected by domain experts to confirm or override the automated gait pattern classification. 
It consists of a Python backend for data management, ML, and explainability as well as a web-based frontend providing the interactive visual interface. 

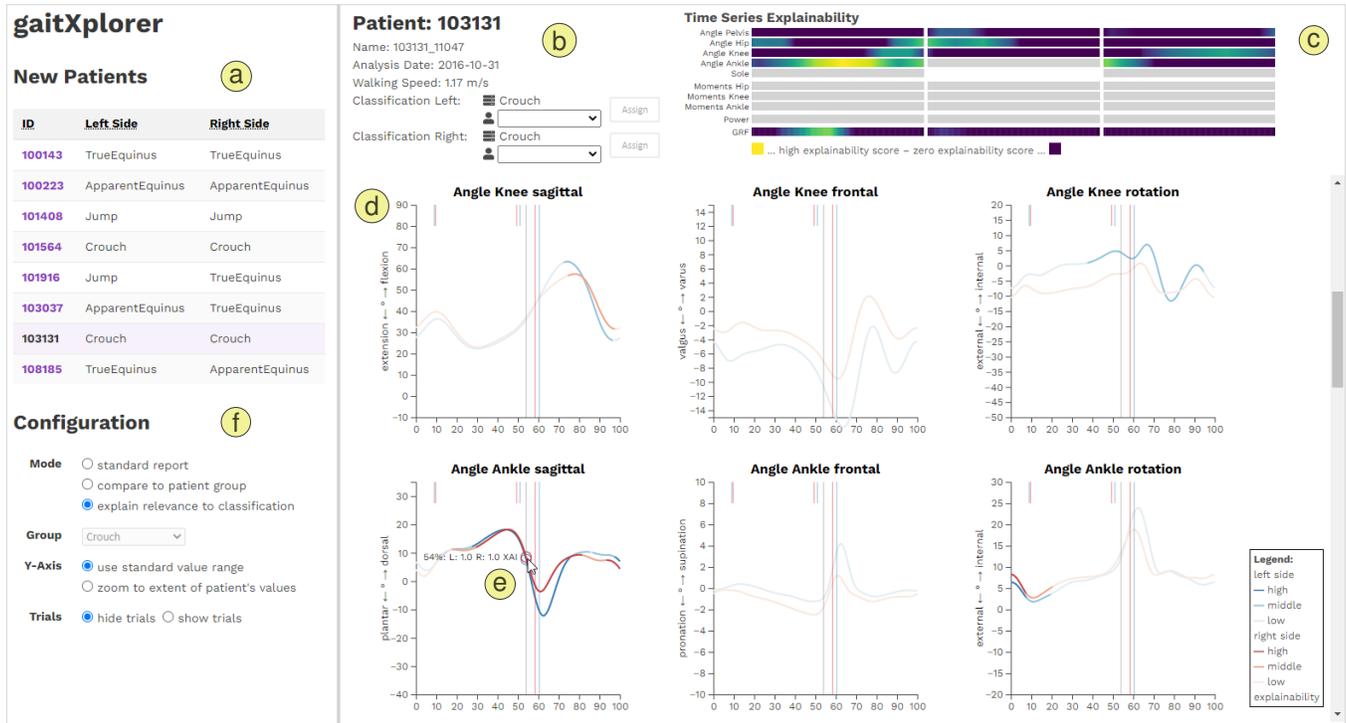
\begin{figure*}[t]
\begin{center}
\begin{tikzpicture}
    \node[anchor=south west,inner sep=0] (image) at (0,0) {\includegraphics[width=1\textwidth]{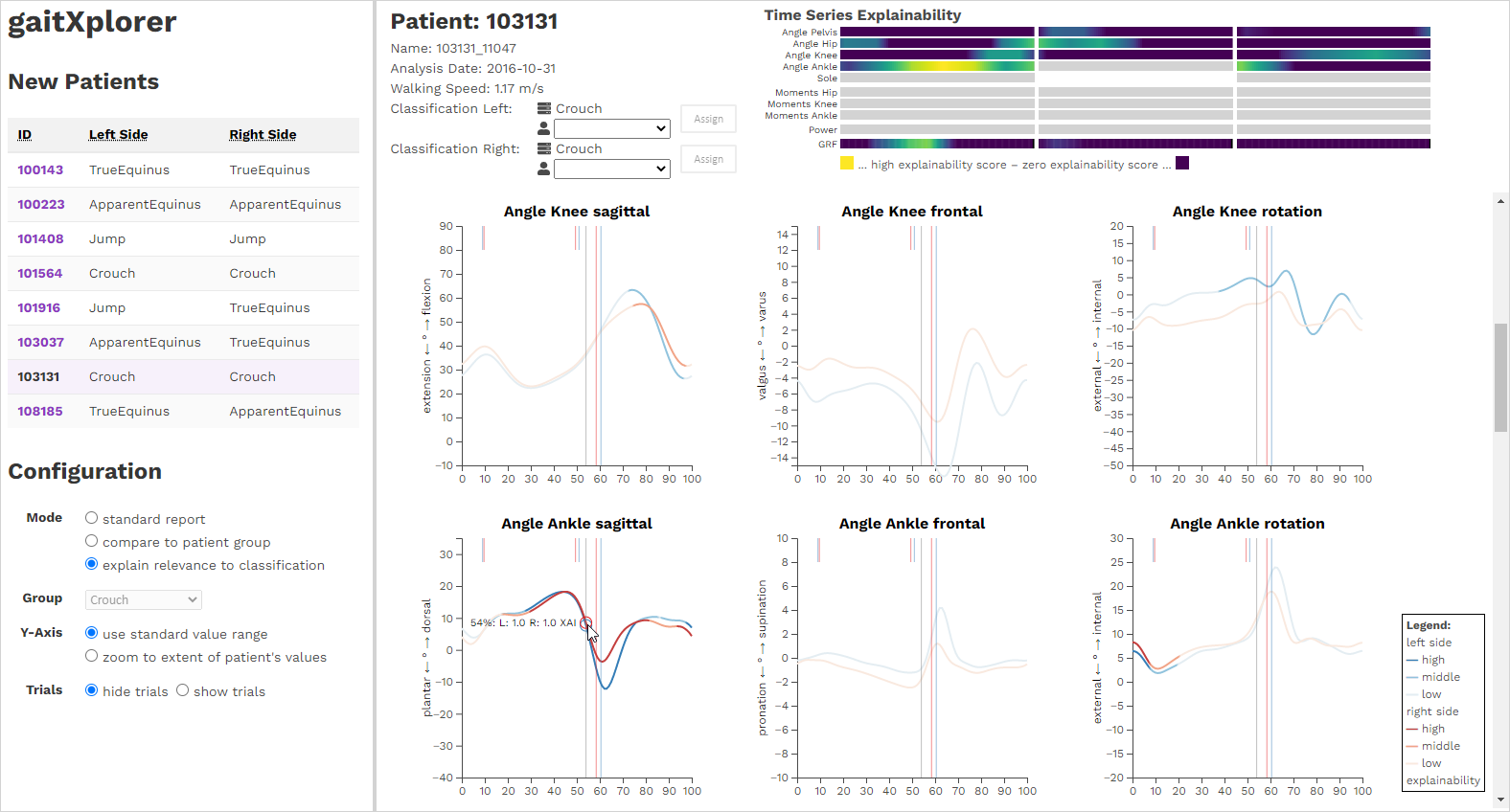}};
    \begin{scope}[x={(image.south east)},y={(image.north west)}]
        \node [image label] at (0.16,0.9) {a};
        \node [image label] at (0.16,0.42) {f};    
        \node [image label] at (0.4,0.95) {b};    
        \node [image label] at (0.965,0.95) {c};    
        \node [image label] at (0.26,0.72) {d};            
        \node [image label] at (0.357,0.195) {e};            
    \end{scope}
\end{tikzpicture}
\end{center}
\caption{Visual interface showing a patient's data in explainability mode: 
(a) list of new patients that need to be classified; 
(b) master data of the current patient including controls to select a gait classification;
(c) compact overview of the patient's time series with color encoding that indicates the relevance (grounding) for the automated prediction;
(d) time series details as line plots with color intensity according to their relevance (discretized into three levels, i.e., high, middle, and low);
(e) cursor displaying the relevance score at the current mouse position; 
(f) controls to switch between modes and change settings.}
\label{fig:uicomplete}
\end{figure*}

\subsection{Data \& Data Management}

We used a retrospective dataset comprising data from 257 children (355 affected legs) with CP, who could walk independently at self-selected walking speed, but had clearly identifiable gait abnormalities. All gait abnormalities were categorized by a clinically well-established procedure~\cite{sangeux_sagittal_2015} into four CP-specific common gait patterns that served as the ground truth, namely: \textbf{true equinus}, \textbf{jump gait}, \textbf{apparent equinus}, and \textbf{crouch gait}.

The data is managed in the Python-based backend, which provides data access to the biomechanical data of the  patients to be classified and to the group averages and standard deviations for each gait classification via a REST API.

\subsection{Explainable Machine Learning}
\label{sec:design-ml}

For the ML pipeline, we used a subset of the available time series that clinical experts mainly focus on, i.e., joint angles in all three anatomical planes of the pelvis, hip, knee, and ankle (only sagittal and transverse plane), as well as all three GRFs. Each of these signals has 101 time points, since they are time-normalized to one gait cycle (0--100\% gait cycle). During a patient's measurement session, multiple gait cycles are recorded. We used the averaged signals per session and body side to account for gait variability within a person. After min-max normalizing each signal we concatenated all signals. Thus, for each ``affected'' leg of a patient we obtained a 1x1414 input vector. The ML model was trained to predict the classes at the level of individual legs. 

As ML method we utilized a Convolutional Neural Network (CNN) model comprised of four consecutive one-dimensional convolutional layers, a flatten layer, a fully-connected layer with 512 neurons, and a softmax output layer with four output neurons. To counteract potential overfitting during training, we used alpha dropout layers prior to the last two fully connected layers with an dropout rate of 0.05. We chose the scaled exponential linear unit (SELU) activation function for all layers. The convolutional layers had the following properties: 64 feature maps, a filter size of three, and a stride of two. We trained the model using the Adam optimizer and the categorical cross-entropy loss function.  

To explain and examine the internal functioning of the trained model we integrated Grad-CAM~\cite{selvaraju2017grad}, a explainability algorithm commonly used for visual data and adapted it to one-dimensional data. Grad-CAM provides explanations for a certain prediction based on the more abstract features learned in the last convolutional layer. The Grad-CAM explanation for an input sample highlights local regions in the input vector that are strongly relevant for the predicted class. This relevance (grounding) for a particular automated prediction is visualized in the interactive visual interface.

The implementation of the ML method and the Grad-CAM method was conducted within the software framework Python~3.7 (Python Software Foundation, USA) and TensorFlow~2.4 (Google Inc., USA).

\subsection{Interactive Visual Interface}

The interactive visual interface (\autoref{fig:uicomplete}) consists of a patient list in the top left, a configuration panel in the bottom left, and patient-specific information in the right, occupying about 75\% of the screen. 

The patient list (\autoref{fig:uicomplete}.a) shows the list of new patients with their automatically predicted gait pattern classification. Clinicians can use the list to navigate between patients.
In the upper center (\autoref{fig:uicomplete}.b), the patient's identifier, the examination date, and their walking speed are shown. 
Below, clinicians can select a gait classification and possibly override the automated classification. 
Note that updating the dataset and the ML model with the newly labeled patients will be in the scope of future work.

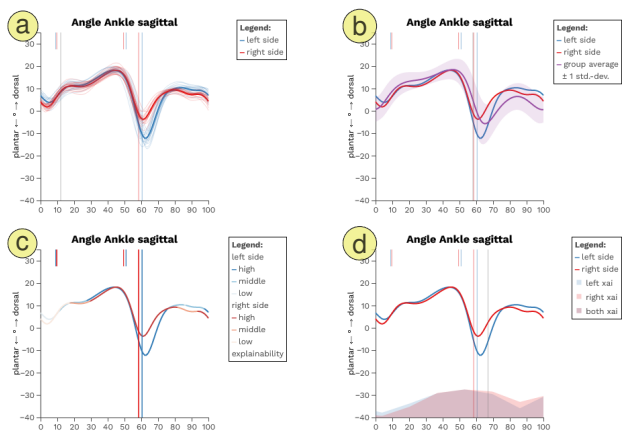
\begin{figure}[ht]
\begin{center}
\begin{tikzpicture}
    \node[anchor=south west,inner sep=0] (image) at (0,0) {\includegraphics[width=0.21\textwidth]{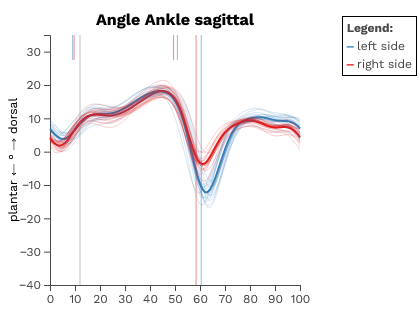}};
    \begin{scope}[x={(image.south east)},y={(image.north west)}]
        \node [image label] at (0.0,0.92) {a};
    \end{scope}
\end{tikzpicture}
\qquad
\begin{tikzpicture}
    \node[anchor=south west,inner sep=0] (image) at (0,0) {\includegraphics[width=0.21\textwidth]{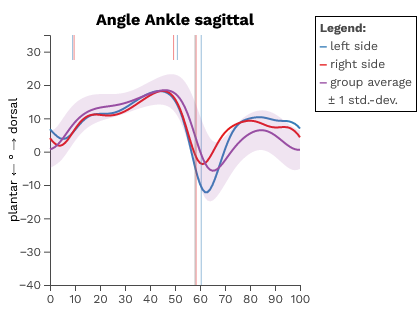}};
    \begin{scope}[x={(image.south east)},y={(image.north west)}]
        \node [image label] at (0.0,0.92) {b};
    \end{scope}
\end{tikzpicture}

\begin{tikzpicture}
    \node[anchor=south west,inner sep=0] (image) at (0,0) {\includegraphics[width=0.21\textwidth]{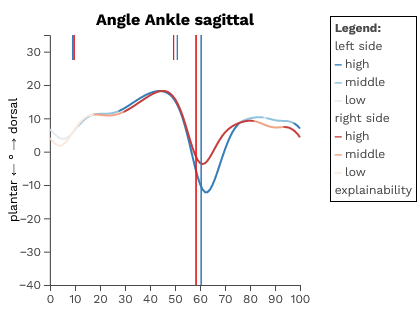}};
    \begin{scope}[x={(image.south east)},y={(image.north west)}]
        \node [image label] at (0.0,0.92) {c};
    \end{scope}
\end{tikzpicture}
\qquad
\begin{tikzpicture}
    \node[anchor=south west,inner sep=0] (image) at (0,0) {\includegraphics[width=0.21\textwidth]{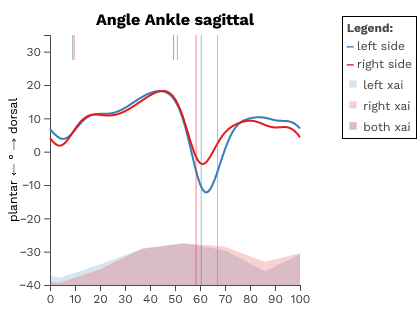}};
    \begin{scope}[x={(image.south east)},y={(image.north west)}]
        \node [image label] at (0.0,0.92) {d};
    \end{scope}
\end{tikzpicture}
\end{center}
\caption{Line plots for detailed inspection of time series (sagittal ankle angles for patient 103131 of the ``\textit{crouch}'' gait pattern): 
(a) standard mode showing the average (thick lines) and all trials (thin lines); 
(b) comparison to the average and standard deviation of all patients with crouch gait in the dataset;
(c) relevance level (high, middle, or low) indicated by color intensity;
(d) relevance scores shown as superimposed area chart in the lower part of the plot.}
\label{fig:lineplots}
\end{figure}

\paragraph{Time series in detail}
The largest part of the screen is used to provide detailed information about the patient's time series (\autoref{fig:uicomplete}.d)
based on the established gait report of our clinical collaborator.
Each left/right pair of time series is displayed in a line plot with the common time axis (x-axis) from 0\% to 100\% gait cycle. 
The y-axis scales are displayed in the physical units of the respective time series (e.g., degree for angles). 
By default, they are scaled to an established value range that includes the values for a broad range of patients. Additionally, the line plots can be zoomed to the value range of the current patient.
The colors blue and red are consistently used for the left and right body side.
As within a patient's measurement session, multiple gait cycles are recorded, by default the averaged time series are visualized. However, it is possible to fade in the individual trials (gait cycles), i.e.\ each step the patient took during the examination (\autoref{fig:lineplots}.a).
Thin vertical support lines indicate events in the gait cycle when the foot and the opposite foot touch (initial contact) and leave the ground (toe-off).
The line plots are arranged in a matrix with three columns for the three anatomical planes and ten rows grouped by biomechanical variable (i.e., joint angles, moments, powers, and GRFs) and body part (e.g., knee, hip, and ankle). 
As stated above, the line plot idiom, the colors used, 
and the matrix arrangement are based on the established gait report 
in order to fit well with the clinicians' work practices. 
We refer to this configuration as the standard mode.

To integrate the annotated dataset and the ML model, we extended the line plots of the standard report with two novel configurations, the explainability mode and the group comparison mode.

The explainability mode (\autoref{fig:lineplots}.c) changes the color intensity of the line plot segments based on the results of the Grad-CAM algorithm. 
For this, the relevance scores are first discretized to three ordinal bins (low $< {}^1{\mskip -5mu/\mskip -3mu}_3$, ${}^1{\mskip -5mu/\mskip -3mu}_3 \leq$ middle  $< {}^2{\mskip -5mu/\mskip -3mu}_3$, high $\geq {}^2{\mskip -5mu/\mskip -3mu}_3$) and then used as color gradient for the line plots. 
In order to maintain the blue/red color convention for body sides and consistent color intensity, colors from ColorBrewer's ``RdBu'' diverging color scheme \cite{d3_chromatic} are used.
We assume that binned relevance levels provide a sufficient level of detail for clinicians and this is in line with design guidelines from social sciences \cite{miller_2019_explanation} that explanation information should be simple. 
In contrast, color gradients from continuous relevance scores are hard to comprehend and compare. 
We also experimented with various multi-hue color schemes from D3 but these made it harder to distinguish left and right side.
Since color encoding of relevance scores was limited to three bins and a spatial encoding can allow for more perceptually effective reading of the explainability information, we prepared an alternative design that represents continuous relevance scores as superimposed area charts in the lower sixth of the line plot (\autoref{fig:lineplots}.d).
Furthermore, an interactive cursor (\autoref{fig:uicomplete}.e) prints the relevance scores for left and right at the current mouse position.

The group comparison mode (\autoref{fig:lineplots}.b) superimposes aggregated information of the patient time series labeled with a certain gait classification. The average is shown as a purple line plot and the range of average plus-minus one standard deviation are shown as a purple band.

\paragraph{Time series overview} 
Since the line plots require space for detailed inspection, it is not possible to show them all at once and scrolling is needed.
To provide a compact overview of all time series and guide the clinician towards interesting parts of the report,
we added a time series heatmap into the top right corner (\autoref{fig:uicomplete}.c). 
Each stripe represents a left/right pair of time series for the 29 relevant combinations of biomechanical variable, body part, and anatomical plane. 
The arrangement of time series is consistent with the matrix of line plots below and the established gait report.
By clicking on a time series stripe, the detailed report scrolls to the corresponding line plot.

\begin{figure}[ht]
\begin{center}
    \includegraphics[width=1\linewidth]{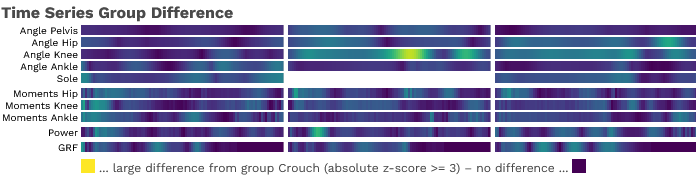}
\end{center}
\caption{Overview plot showing group differences (comparing patient 103131 to the crouch gait class).}
\label{fig:overview-group}
\end{figure}

While the line plots show two time series for both legs, each stripe of the compact overview plot displays a single time series that should represent how interesting this line plot is.
The calculation of this aggregated time series depends on the report mode:
In the explainability mode, the overview plot presents the maximum of the left and right relevance score for each time point (\autoref{fig:uicomplete}.c).
In the group comparison mode, the interesting time points have a large difference to the group average relative to the group's standard deviation. 
We calculate the maximum of the absolute values of the z-scores for the left and right side at each time point (\autoref{fig:overview-group}).
In the standard mode, asymmetry between the legs is assumed to be interesting~\cite{wagner_2018_kavagait}, which we quantify as the difference between left and right relative to the total extent of values in this stripe.

The heatmap uses the ``viridis'' sequential color scheme by van der Walt, Smith, and Firing \cite{d3_chromatic} for all modes.
An alternative design with distinct color schemes for each mode was shown and discussed in the case study.

In the bottom left part is a configuration panel (\autoref{fig:uicomplete}.f) to switch modes, select a patient group for comparison, zoom the y-axis, and show individual trials, i.e. each step recorded during the examination.
The visual interface was implemented using TypeScript 4.3, D3 7.0, Svelte 3.49, Bulma 0.9, and Font Awesome 5.15 icons. 

\section{Case Study}
\label{sec:eval}
We investigated the applicability of this visual analytics approach in a case study based on annotated patient data from our collaboration partner and collected their feedback in a series of focus group interviews. 
The case study builds upon a long-term, iterative collaboration between the involved clinical experts and the researchers, especially in the field of ML. 
The two clinical experts, who are also coauthors of this work, have formal education in sport science and training and over 29 years and seven years, respectively, of practical experience in clinical gait analysis.

As described in \autoref{sec:design-ml}, the ML model was trained on a dataset comprising data from 257 children with CP (355 affected legs).
To simulate new unlabeled data for demonstration in the focus groups, we drew a random subsample of eight patients so that two samples were available for each class (only the left legs were considered during the selection process). In addition, for this subsample we selected only samples that were correctly classified by the ML model. This information was of course not shared with the clinicians in advance.

For the design iteration described in the work at hand, we conducted three focus group interviews of about one hour over the video conference platform MS Teams. 
We demonstrated the visual interface over MS Teams' screen sharing feature based on the pair analytics method \cite{ariashernandez_2011_pair}. The clinical experts gave spoken inputs to navigate in the data visualization and the facilitator (i.e., the visual analytics expert) interacted with the tool accordingly.
In these interviews we discussed the usability of the visual interface and how such a visual analytics tool can be integrated into the clinical workflow.
In addition, the clinical experts received screenshots of the eight patients showing all data visualizations (cp.\ supplemental material) 
and were asked whether these explanations made a plausible or suspicious impression.
We also discussed how the visual interface affects their trust in the automated classifications.

\paragraph{Visual Interface Design}
The clinicians found the visual interface clear and intuitive. 
They appreciated the reuse of the line plot style and matrix arrangement from their established layout.

Showing the explainability results in ordinal bins of high, middle, and low relevance level was regarded as sufficient and the color gradients were confirmed as readable. 
The clinicians emphasized the need for a color legend, which needs to clarify in particular how regions with a zero relevance score are handled.
They pointed out that the color gradients are not readable when the two line plots for the left and right side overlap (e.g., \autoref{fig:lineplots}.c).
The interactive cursor showing the relevance score at the mouse position served as a workaround for such regions.
The alternative design using area plots for relevance scores (\autoref{fig:lineplots}.d) made a good first impression but on closer inspection the superimposed area plots with areas of mixed colors were hard to interpret. 
While discussing various further alternatives the clinicians expressed their preference for the color gradient rather than adding more visual elements to the screen. 

A stable color scheme for the compact time series overview was preferred by the clinical experts over three color schemes for each mode.
The clinicians expressed that familiarizing with different color schemes poses more burden than the concern of confusing the current mode. In addition, the selected mode can be easily recognized from the detailed line plots.
They asked for the stripes to be larger, so that they would be easier to discern and click on. 
Thus, after the first interview the stripe height was increased from 5 to 10 pixels and tiny labels determining the biomechanical variable and body part were added.
Six rows of time series, angle speeds and angle acceleration, were removed because they are not needed for analysis of patients with CP.

\paragraph{Explanations \& Trust}
The two clinical experts independently inspected data visualizations of eight patients and assessed their trust in the automated classification based on the displayed relevance scores. 
All the visualizations can be found in the supplemental material and can be looked up using the 6-digit patient identifier.

Over the eight patients and 16 legs, the clinicians stated they would rather trust the explanations for both legs of two patients (101408, 101564) and for one leg of two patients (101916 left, 108185 right). For one patient the clinicians gave diverging answers (100223). In this case, both clinicians considered the relevant regions to be plausible, but one of the clinicians expected higher relevant regions in the sagittal knee angle. For the remaining legs (100143, 103037, 103131, 101916 right, 108185 left), the clinicians were distrusting the explanations (\autoref{tab:plausible}).

\begin{table}[th]
  \caption{Plausibility assessment of explanations.}
  \label{tab:plausible}
  \centering
\begin{tabular}{l l l}
\toprule
patient id & left & right \\
\midrule
100143     & suspicious & suspicious  \\
100223     & diverging   & diverging     \\
101408     & plausible  & plausible   \\
101564     & plausible  & plausible   \\
101916     & plausible  & suspicious  \\
103037     & suspicious & suspicious  \\
103131     & suspicious & suspicious  \\
108185     & suspicious & plausible  \\
\bottomrule
\end{tabular}

\end{table}

Summarizing the feedback, there were two types of suspicious patterns in the explanations: 
On the one hand, certain time series regions had a lower relevance score than the clinicians expected.
For example, patient 103131's gait (\autoref{fig:uicomplete}) is classified as crouch gait which has a characteristic behavior in the sagittal knee angle, but that time series contains only low to middle relevance scores. 
On the other hand, the relevance score was high for time series regions that the clinicians did not regard as characteristic for the predicted class. 
Low relevance scores were more frequently criticized. 
Often too low relevance scores in some regions (e.g., sagittal knee or ankle angles) were accompanied by too high relevance scores in other (for clinicians unexpected) regions. 
Samples classified as true equinus exhibited often this suspicious pattern, e.g., both legs of patient 100143, classified as true equinus, had high relevance scores for the expected frontal knee angle but low relevance scores for the sagittal ankle angle. The clinicians expected relevant regions in the sagittal ankle angle, as this is the most important biomechanical variable for characterizing true equinus. However, for the true equinus class, the ML model identified greater differences in sagittal knee angle compared to the other CP-related gait patterns, which were sufficient for classification.

For all legs classified as jump gait (101408 both, 101916 left), the explanations were rated as plausible. This is notable as explanations for this class contained much larger regions of high relevance scores than explanations for other classes. Even though this is clearly visible in the compact time series overview plots with more yellow\slash{}green regions, the clinicians did not consider this as  a sign for distrust.

The clinicians expressed certain concerns about the ML model. In summary, confidence in the ML model was more affected by missing relevance scores for regions where they expected them than for regions that had high relevance scores but were not expected.  Additionally, they emphasized that a visual interface that makes the algorithm more transparent and provides such detailed insights into its functioning is essential to gain trust in the use of ML-based classification in the clinical setting.

\paragraph{Integration in Workflows}
The clinicians stated that the group comparison mode is good and an interesting addition to their standard report.
However, the most important addition was the inclusion of explainability and visualization thereof, which enabled the evaluation of the functioning of the automatic classification algorithm. 
The clinicians clearly stated that this should be a part of every clinical gait analysis that utilizes ML in order to achieve the best possible treatment outcome for the patient.

Reflecting on the analysis of the patient visualizations (cp.\ supplemental material), the senior clinician described his workflow as follows:
First, he looked at the detailed line plots from top to bottom. 
Normally, he would use their standard report; here he took the explainability mode ignoring color gradients.
Based on the time series data, he decided on a CP-related gait pattern.
Finally, he looked at the relevance scores and compared them to his decision and expectations.
He did not consult the group comparison mode because he is familiar with the time series for the different gait groups in the dataset.
He neither used the compact time series overview, because he always reads a patient report from top to bottom.

\section{Discussion \& Future Work}
\label{sec:discuss}

An interesting observation from the focus group was that there was a certain discrepancy between clinicians' expectations and the signal regions actually used by the ML model to make its predictions (i.e., the visualized relevance scores). The clinicians expected the ML model to use all regions unique to the specific class (and thus regions where the samples of one class differ from the other classes), but actually the ML model seemed to use only a subset of all these regions. This is evident in the samples classified as true equinus, where the model used regions in the sagittal hip and knee angles, but the clinicians expected the model to use primarily the sagittal ankle angle, as this is considered  the main biomechanical variable for this gait pattern.  
The expectation that relevant regions include all clinically relevant variables may also be fostered by clinicians' experience with methods such as statistical parametric mapping (SPM)~\cite{pataky_generalized_2010}, a method to identify statistically significant differences between two different patient groups, but which cannot be used for automated classification.

The explainability-enriched visual analytics approach still has limitations.
First, there is no feedback mechanism that would allow the clinicians to externalize their domain knowledge on relevant regions and thus influence the training of the ML model. 
A natural evolution of this work will be to integrate such feedback into future iterations of this approach. 
Second, the Grad-CAM algorithm determines its explanations solely from the relevance of input data to the predicted classification. 
There is a broad landscape of alternative explanation approaches~\cite{adadi2018peeking}.
For example, counterfactual explanations could point out how much a gait pattern would need to change until it is classified differently.
It will be a fruitful area for further work to investigate how clinicians react to counterfactual explanations or a combination with relevance explanations.
Third, most patients in clinical practice suffer from two or more gait abnormalities affecting also the frontal and transversal plane (e.g., true equinus and valgus deformity). 
To incorporate overlapping gait abnormalities, a multi-label ML approach and set-visualization techniques can be incorporated.
Finally, the case study evaluation has the limitation that the two domain experts were already involved in the conception of the visual analytics approach.
Therefore, studies with additional clinicians are needed in the future.
However, to broaden the user base, onboarding mechanisms \cite{stoiber_2019_visualizationb} need to be provided. These mechanisms would support clinical experts in learning to use the visual analytics tool and to extract information from the visualizations. 

\section{Conclusion}
\label{sec:conclusion}
In this work, we present  \textit{gaitXplorer}, a visual analytics approach for clinical gait analysis of patients with CP that we developed and evaluated with two clinical experts. 
While we focus on CP as a case study, the visual analytics approach is applicable to clinical gait analysis in general, given that labeled training data is available.
Even beyond the clinical context, the results can be generalized to other visual analytics scenarios with multiple interrelated time series. Our approach can serve as a reference for the integration of an explainability algorithm into a ML-based visual analytics approach. 
Overall, this case study strengthens the idea that visual analytics approaches, which integrate an explainability algorithm and embed its explanations into the interactive visual interface, are essential to gain confidence in the use of ML-based classification in clinical settings.

\quad \newline 
\acknowledgments{
This work was partly funded
by the Austrian Research Promotion Agency (FFG, \#866855), 
by the Austrian Science Fund (FWF): P33531-N, 
as well as by the Gesellschaft für Forschungsförderung NÖ (Research Promotion Agency of Lower Austria) and the Provincial Government of Lower Austria within IntelliGait3D (\#FTI17-014) and within the Endowed Professorship for Applied Biomechanics and Rehabilitation Research (\#SP19-004).
}

\bibliographystyle{abbrv-doi-hyperref-narrow}

\bibliography{remocaplab}
\end{document}


\firstsection{Description}

\maketitle

This document contains screenshots of the visual interface for a sample of eight patients. 
For the purpose of screenshots, the left part of the user interface and the scrolling were removed. 
The line plot legend was moved from the bottom right to the top-most line plot.

Each figure depicts the patient in group comparison mode on the left and in explainability mode on the right.
Please zoom in to view the screenshots in their original resolution and consult the manuscript for a description of the design.

\begin{figure*}[p]
\begin{center}
    \includegraphics[width=0.42\linewidth]{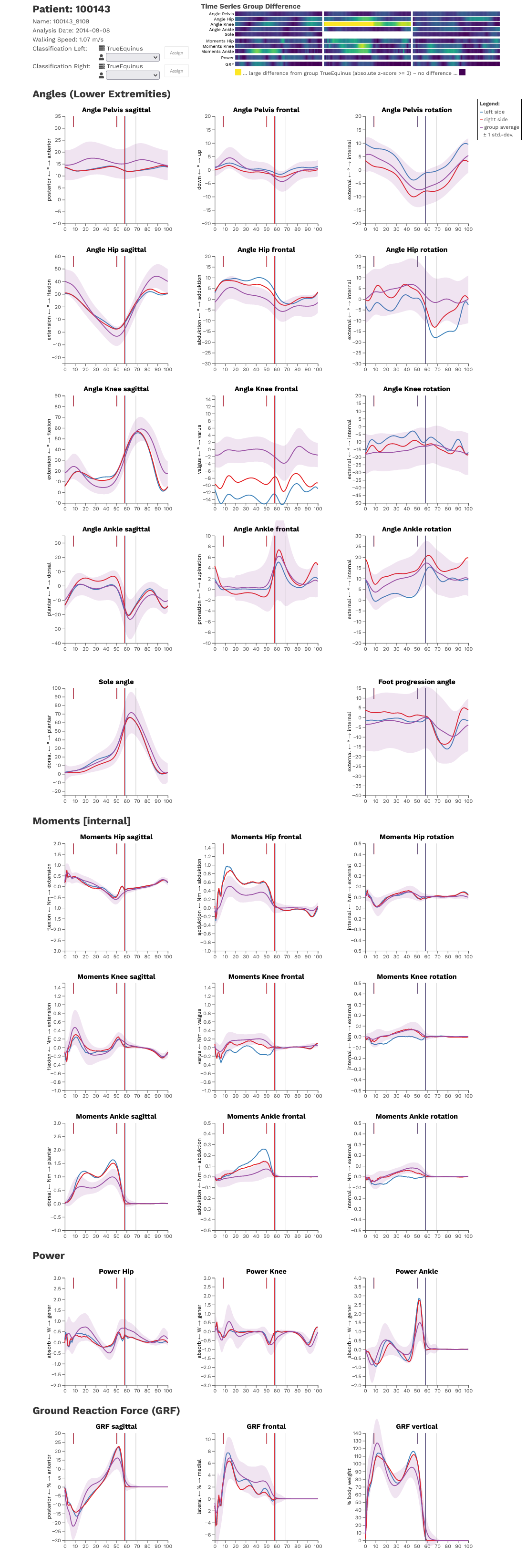}
    \quad
    \includegraphics[width=0.42\linewidth]{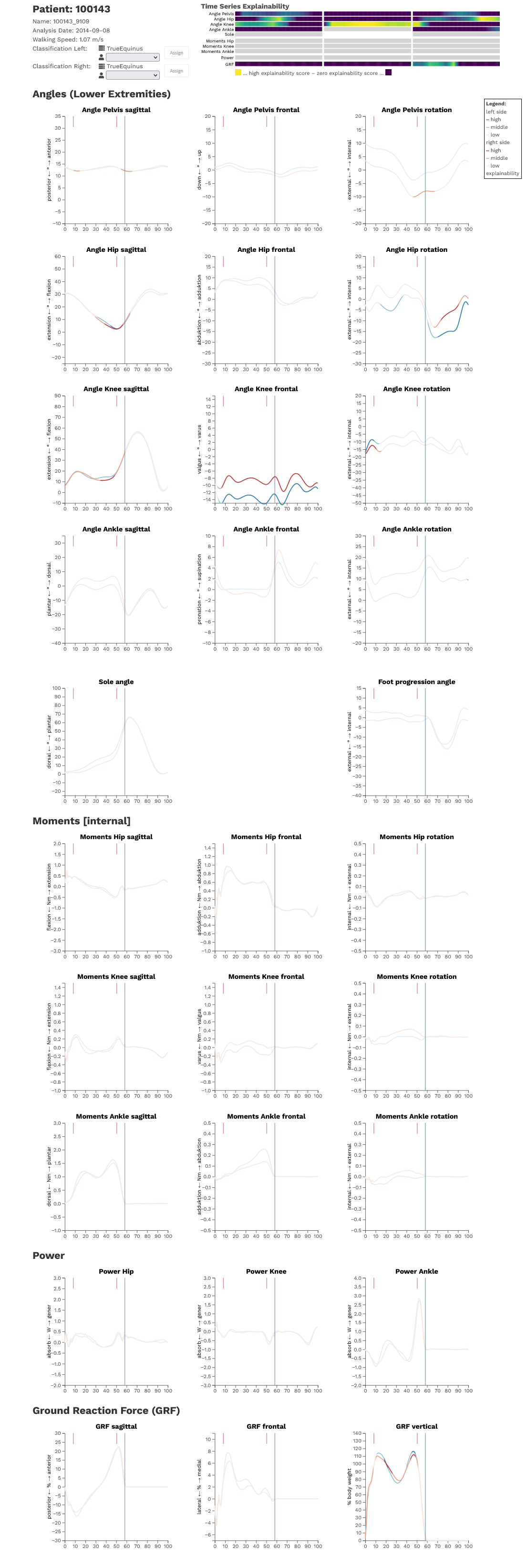}
\end{center}
\caption{Patient 100143 (left: TrueEquinus, right: TrueEquinus, group: TrueEquinus)}
\label{fig:p100143}
\end{figure*}

\begin{figure*}[p]
\begin{center}
    \includegraphics[width=0.42\linewidth]{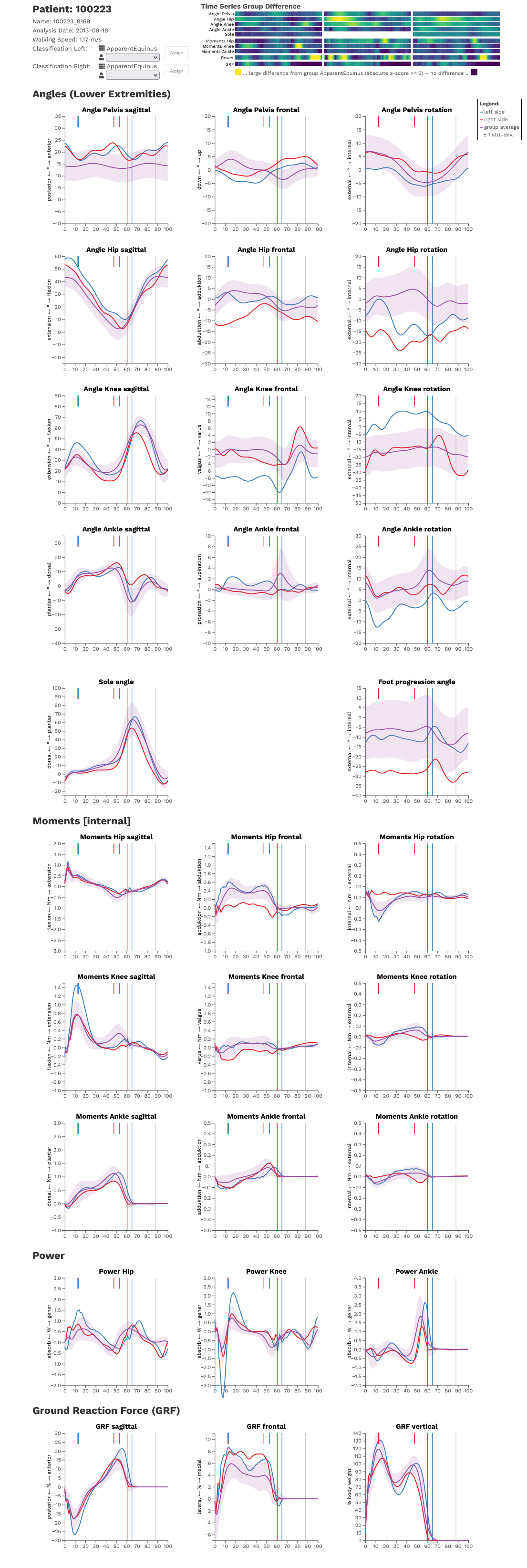}
    \quad
    \includegraphics[width=0.42\linewidth]{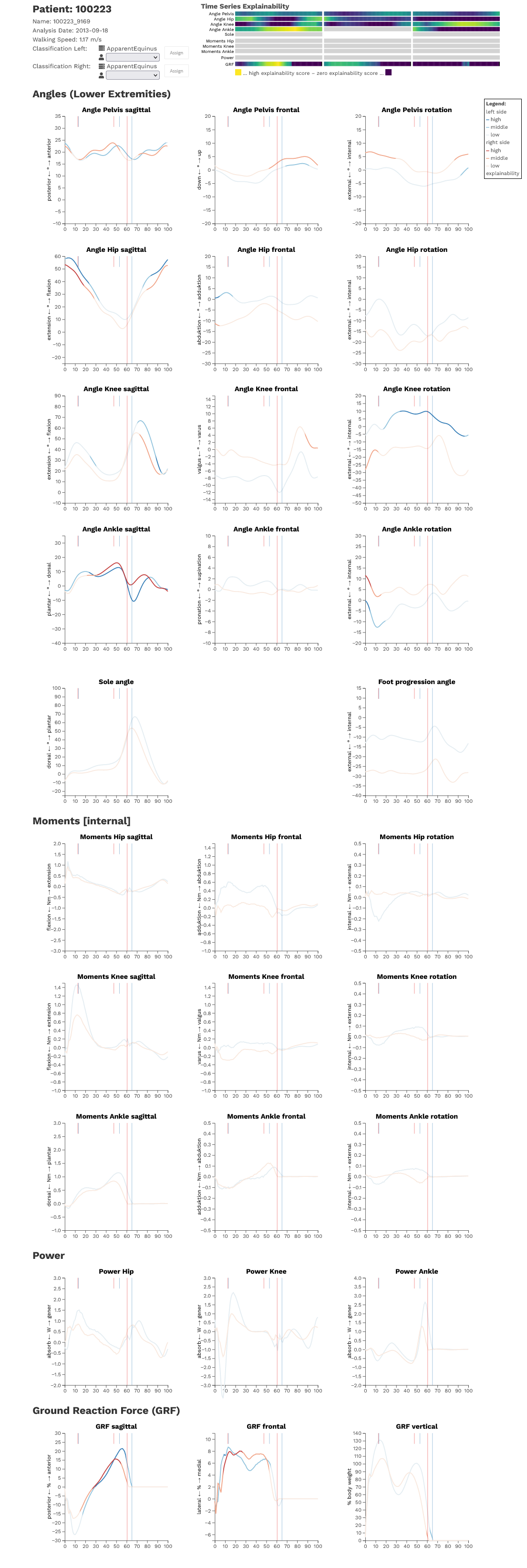}
\end{center}
\caption{Patient 100223 (left: ApparentEquinus, right: ApparentEquinus, group: ApparentEquinus)}
\label{fig:p100223}
\end{figure*}

\begin{figure*}[p]
\begin{center}
    \includegraphics[width=0.42\linewidth]{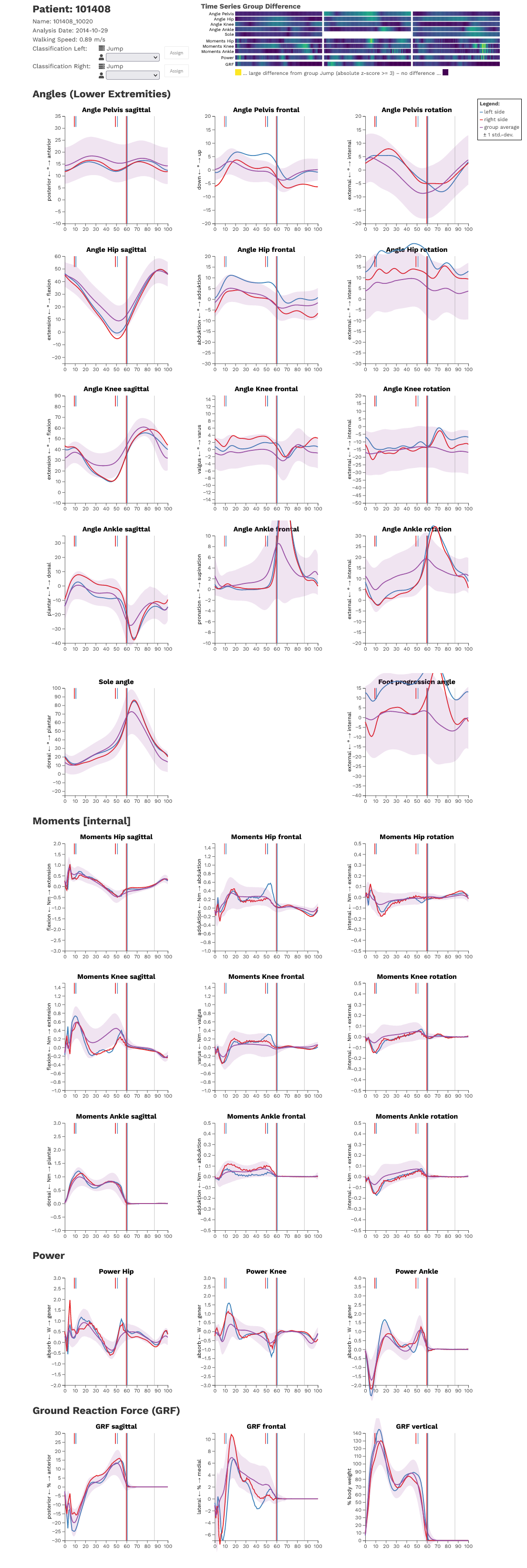}
    \quad
    \includegraphics[width=0.42\linewidth]{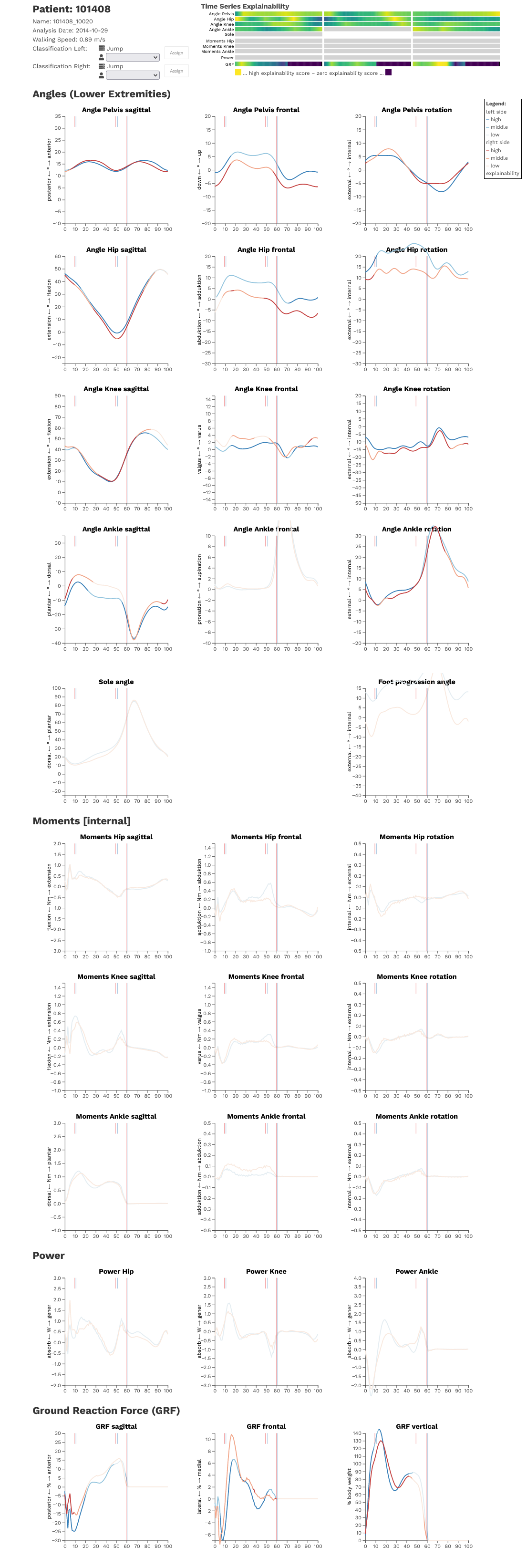}
\end{center}
\caption{Patient 101408 (left: Jump, right: Jump, group: Jump)}
\label{fig:p101408}
\end{figure*}

\begin{figure*}[p]
\begin{center}
    \includegraphics[width=0.42\linewidth]{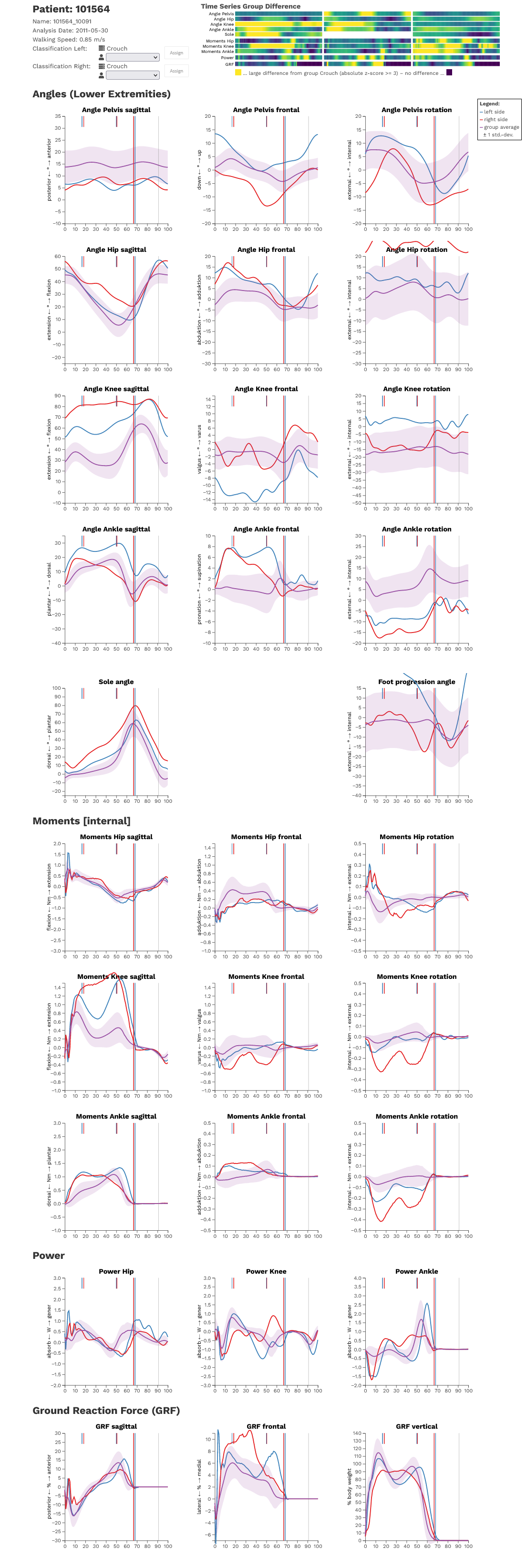}
    \quad
    \includegraphics[width=0.42\linewidth]{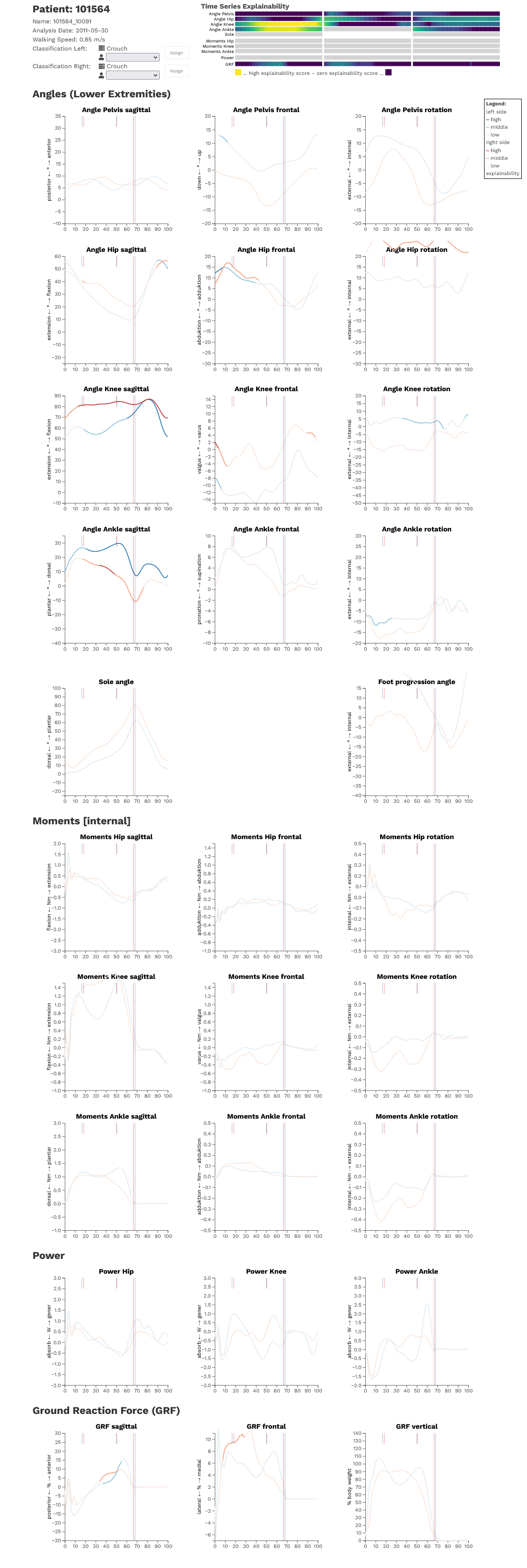}
\end{center}
\caption{Patient 101564 (left: Crouch, right: Crouch, group: Crouch)}
\label{fig:p101564}
\end{figure*}

\begin{figure*}[p]
\begin{center}
    \includegraphics[width=0.42\linewidth]{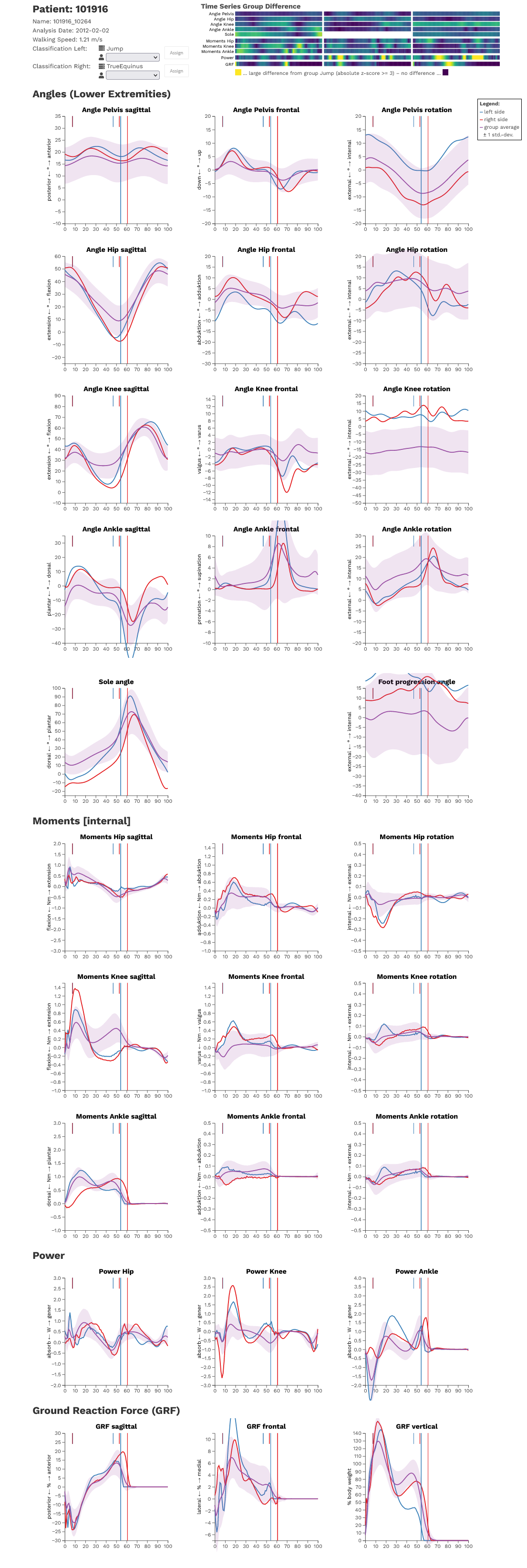}
    \quad
    \includegraphics[width=0.42\linewidth]{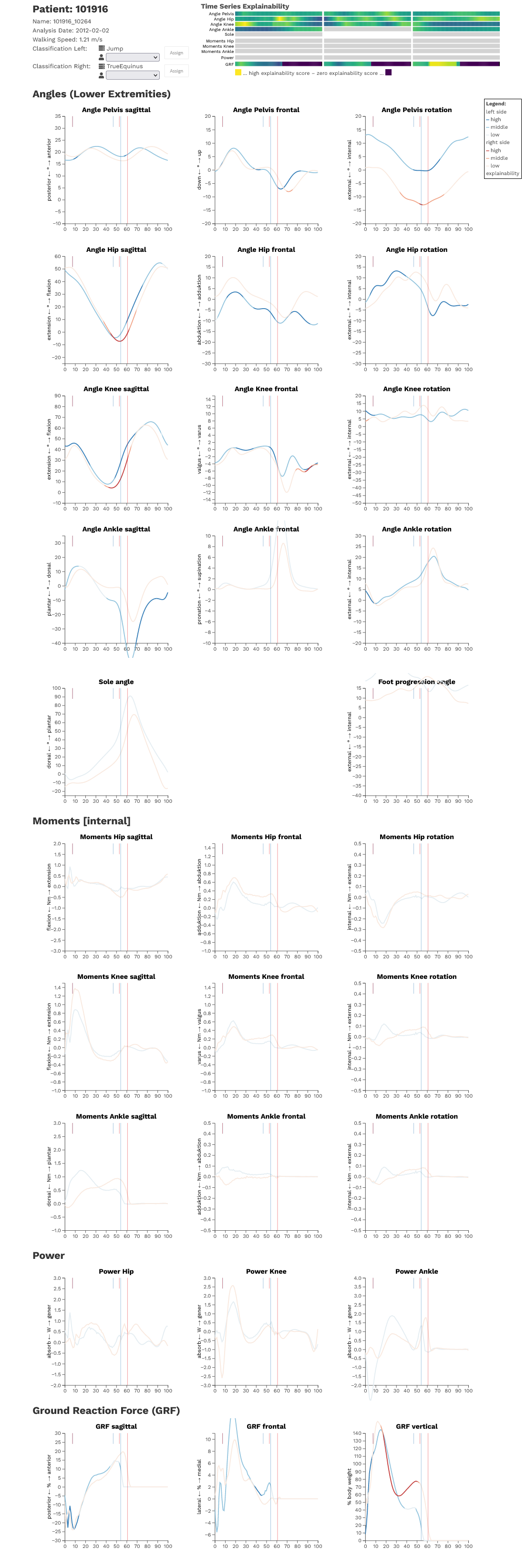}
\end{center}
\caption{Patient 101916 (left: Jump, right: TrueEquinus, group: Jump)}
\label{fig:p101916}
\end{figure*}

\begin{figure*}[p]
\begin{center}
    \includegraphics[width=0.42\linewidth]{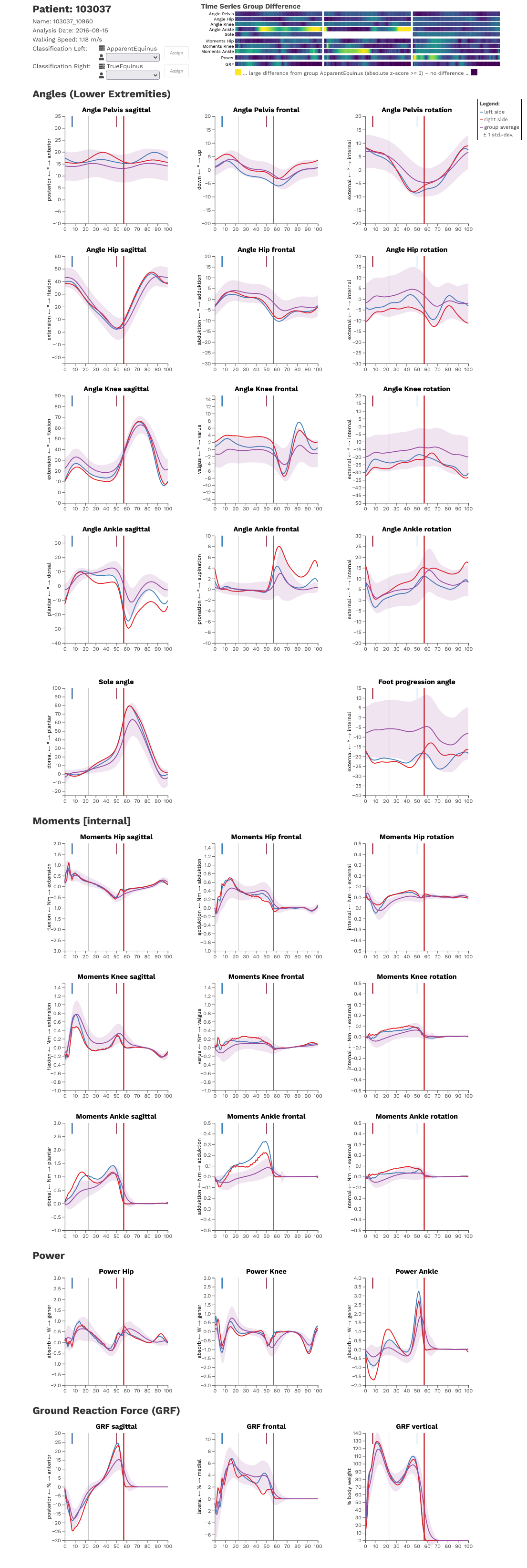}
    \quad
    \includegraphics[width=0.42\linewidth]{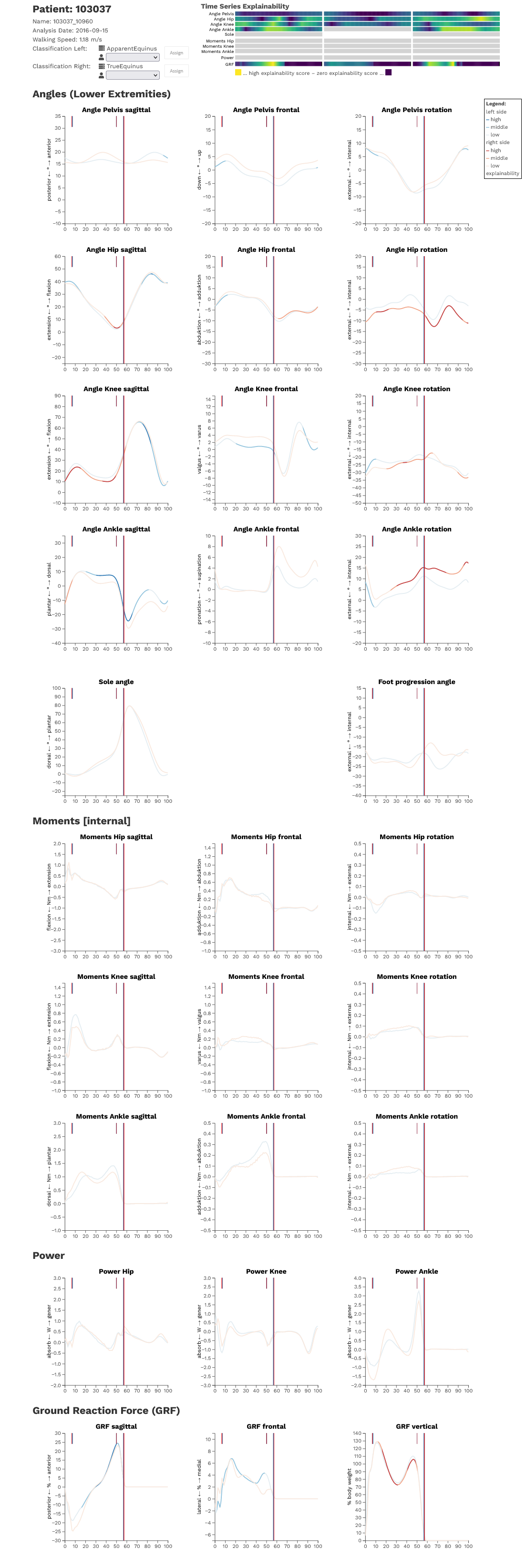}
\end{center}
\caption{Patient 103037 (left: ApparentEquinus, right : TrueEquinus, group: ApparentEquinus)}
\label{fig:p103037}
\end{figure*}

\begin{figure*}[p]
\begin{center}
    \includegraphics[width=0.42\linewidth]{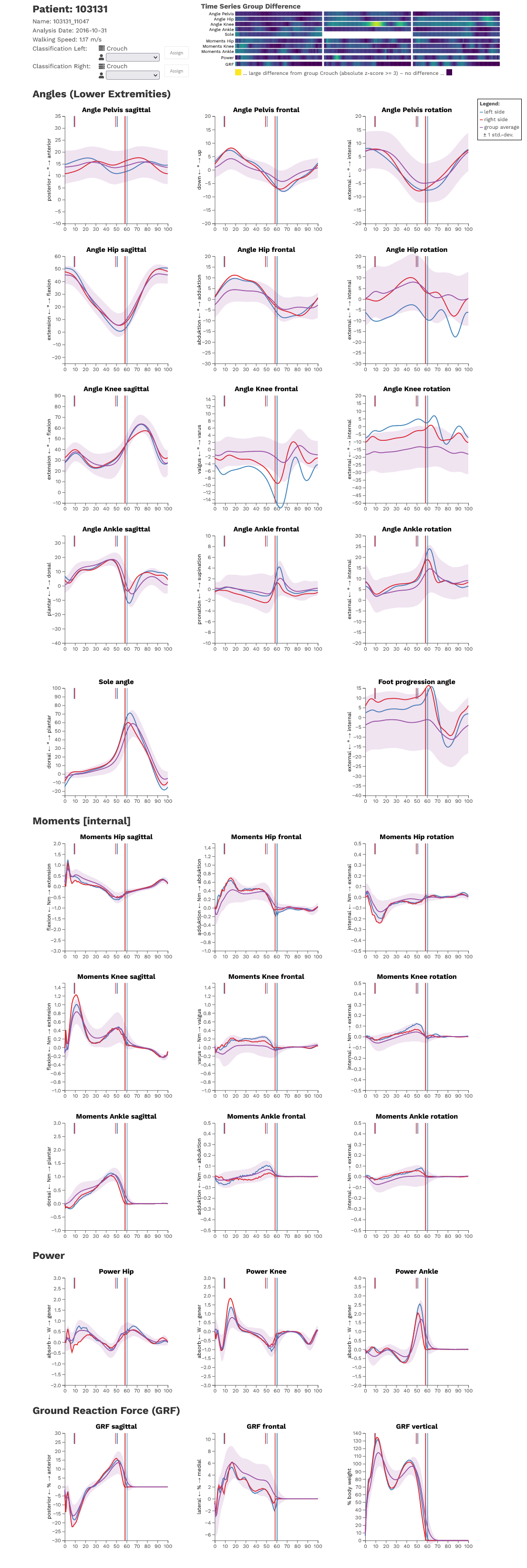}
    \quad
    \includegraphics[width=0.42\linewidth]{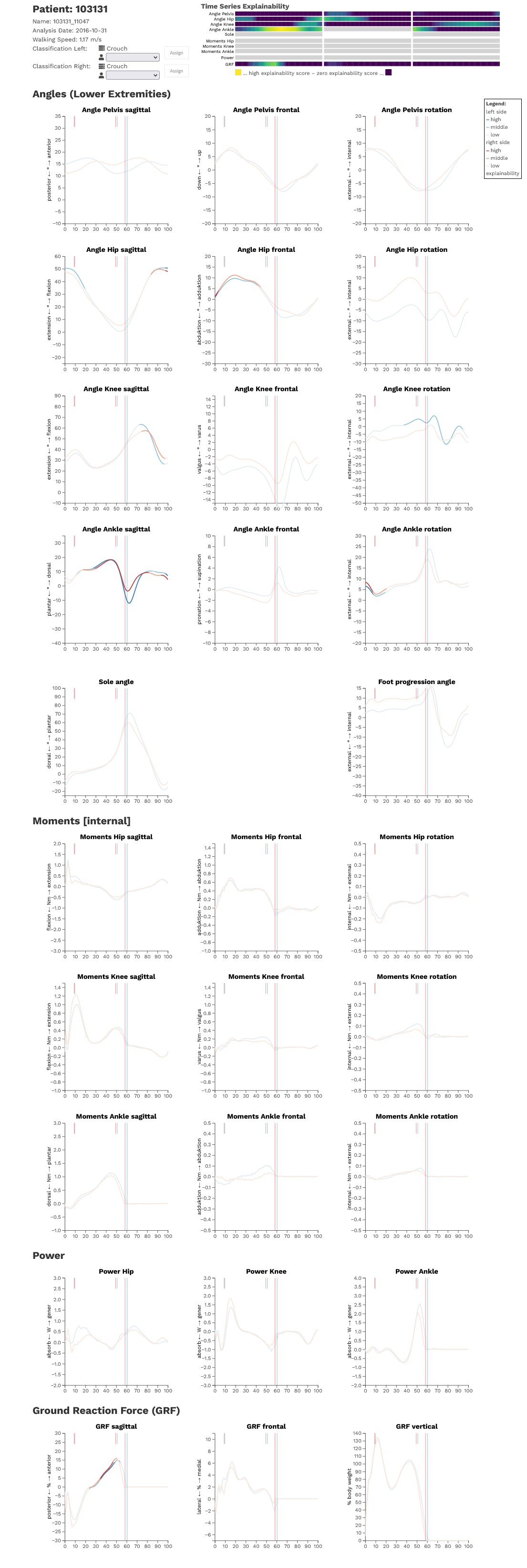}
\end{center}
\caption{Patient 103131 (left: Crouch, right: Crouch, group: Crouch)}
\label{fig:p103131}
\end{figure*}

\begin{figure*}[p]
\begin{center}
    \includegraphics[width=0.42\linewidth]{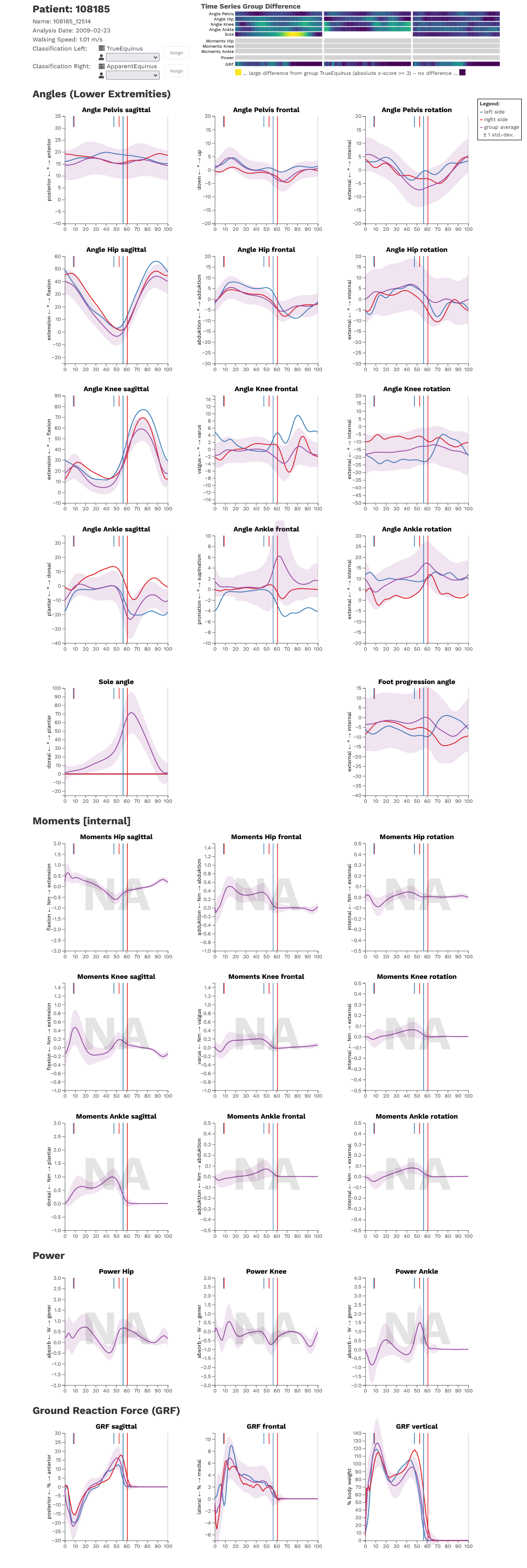}
    \quad
    \includegraphics[width=0.42\linewidth]{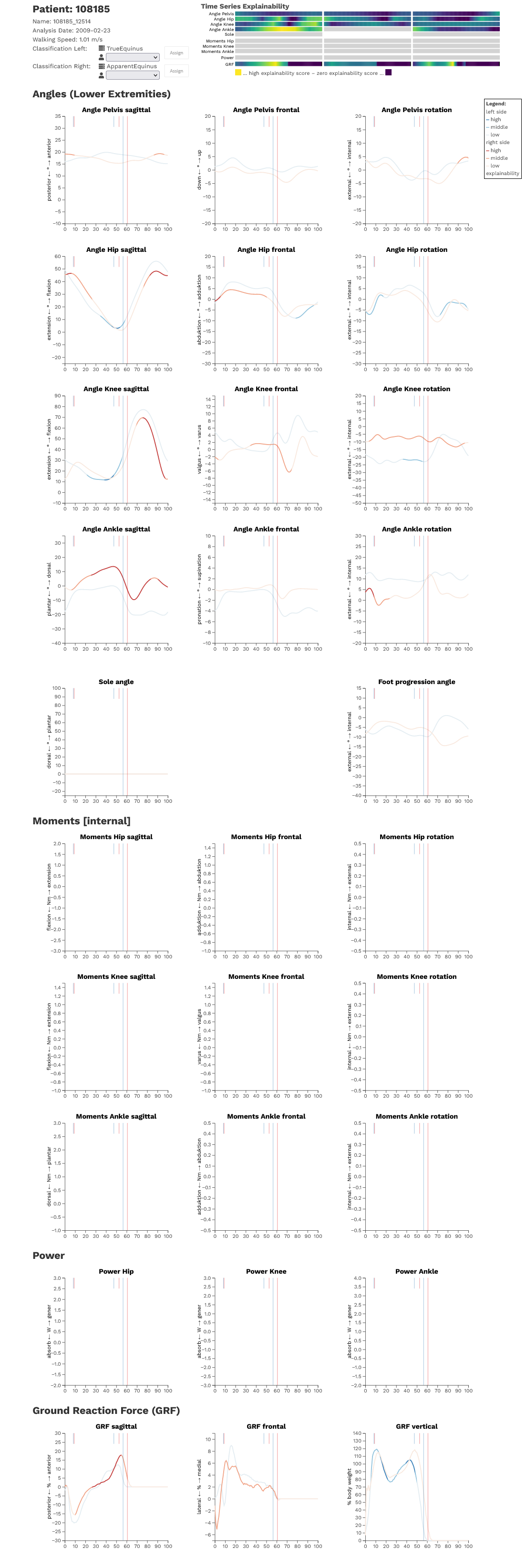}
\end{center}
\caption{Patient 108185 (left: TrueEquinus, right: ApparentEquinus, group: TrueEquinus)}
\label{fig:p108185}
\end{figure*}

\acknowledgments{
This work was partly funded
by the Austrian Research Promotion Agency (FFG, \#866855), 
by the Austrian Science Fund (FWF): P33531-N, 
as well as by the Gesellschaft für Forschungsförderung NÖ (Research Promotion Agency of Lower Austria) and the Provincial Government of Lower Austria within IntelliGait3D (\#FTI17-014) and within the Endowed Professorship for Applied Biomechanics and Rehabilitation Research (\#SP19-004).
}